\documentclass[journal,twocolumn,twoside]{IEEEtran}

\ifCLASSINFOpdf
\usepackage[pdftex]{graphicx}
\else
\usepackage[dvips]{graphicx}
\fi

\usepackage[cmex10]{amsmath}
\usepackage{amsfonts,amsbsy,amssymb,amstext,mathrsfs,latexsym,color,url}

\newtheorem{theorem}{Theorem}[section]
\newtheorem{lemma}[theorem]{Lemma}
\newtheorem{proposition}[theorem]{Proposition}
\newtheorem{corollary}[theorem]{Corollary}

\newtheorem{definition}[theorem]{Definition}

\newcommand{\bitem}{\begin{itemize}}
\newcommand{\eitem}{\end{itemize}}
\newcommand{\beq}{\begin{equation}}
\newcommand{\eeq}{\end{equation}}
\def\supp{{\text{\rm supp}}}
\def\tr{{\text{\rm tr}}}
\def\span{{\text{\rm span}}}

\def\argmin{{\text{\rm argmin}} \,}

\newcommand{\cW}{{\mathcal W}}
\newcommand{\cK}{{\mathcal K}}

\newcommand{\RR}{{\mathbb R}}
\newcommand{\EE}{{\mathbb E}}

\newcommand{\PP}{{\mathbb P}}

\def \x{\mathbf{x}}

\def \y{\mathbf{y}}
\def \u{\mathbf{u}}
\def \vv{\mathbf{v}}
\def \h{\mathbf{h}}

\def \z{\mathbf{z}}

\def \b{\mathbf{b}}

\def \a{\mathbf{a}}

\def \c{\mathbf{c}}

\def \H{\mathbf{H}}

\def \U{\mathbf{U}}

\def \I{\mathbf{I}}

\def \A{\mathbf{A}}

\def \P{\mathbf{P}}

\def \X{\mathbf{X}}
\def \Y{\mathbf{Y}}
\def \Z{\mathbf{Z}}

\def \0{\mathbf{0}}
\def \tr{\textnormal{tr}}

\def \tt {\widetilde{t}}

\def\cH{\mathcal{H}}

\def\cW{\mathcal{W}}

\newcommand{\ip}[2]{\left\langle#1,#2\right\rangle}

\newcommand{\absip}[2]{| \langle#1,#2\rangle |}
\newcommand{\norm}[1]{\|#1\|}

\begin{document}

\title{Sparse Recovery from Combined Fusion Frame Measurements}

\author{Petros~Boufounos,~\IEEEmembership{Member,~IEEE,}
        Gitta~Kutyniok,
        and~Holger~Rauhut%
    \thanks{ Manuscript received December 10, 2009; revised November
      16, 2010; accepted December 14, 2010. Date of current version
      May 25, 2011. P. Boufounos did part of this work while at Rice
      university. P. Boufounos was supported in part by the grants NSF
      CCF-0431150, CCF-0728867, CNS-0435425, and CNS-0520280,
      DARPA/ONR N66001-08-1-2065, ONR N00014-07-1-0936,
      N00014-08-1-1067, N00014-08-1-1112, and N00014-08-1-1066, AFOSR
      FA9550-07-1-0301, ARO MURI W311NF-07-1-0185, and the Texas
      Instruments Leadership University Program. P. Boufounos was
      supported in part by Mitsubishi Electric Research Laboratories
      (MERL). G. Kutyniok was in part supported by Deutsche
      Forschungsgemeinschaft (DFG) Heisenberg Fellowship KU 1446/8,
      DFG Grant SPP-1324, KU 1446/13, and DFG Grant KU
      1446/14. H. Rauhut acknowledges support by the Hausdorff Center
      for Mathematics and by the WWTF project SPORTS (MA 07-004).
}
        \thanks{P. Boufounos is with Mitsubishi Electric Research
  Laboratories, {\em petrosb@merl.com}.}
\thanks{G. Kutyniok is with Institute of Mathematics, University of
  Osnabr\"uck, {\em kutyniok@uni-osnabrueck.de}.}
\thanks{H. Rauhut is with Hausdorff Center for Mathematics \&
  Institute for Numerical Simulation, University of Bonn, {\em
    rauhut@hcm.uni-bonn.de}.}%
\thanks{Communicated by J. Romberg, Associate Editor for Signal
  Processing.}%
\thanks{Digital Object Identifier 10.1109/TIT.2011.2143890}}%

\markboth{IEEE Transaction on Information Theory, VOL. 57, NO. 6, JUNE 2011}%
{Boufounos \lowercase{{\em et al.}}: Sparse Recovery from Combined Fusion Frame Measurements}

\maketitle

\begin{abstract}
Sparse representations have emerged as a powerful tool in signal and
information processing, culminated by the success of new acquisition
and processing techniques such as Compressed Sensing (CS). Fusion
frames are very rich new signal representation methods that use
collections of subspaces instead of vectors to represent signals. This
work combines these exciting fields to introduce a new sparsity model
for fusion frames. Signals that are sparse under the new model can be
compressively sampled and uniquely reconstructed in ways similar to
sparse signals using standard CS. The combination provides a promising
new set of mathematical tools and signal models useful in a variety of
applications. With the new model, a sparse signal has energy in very
few of the subspaces of the fusion frame, although it does not need to
be sparse within each of the subspaces it occupies. This sparsity
model is captured using a mixed $\ell_1/\ell_2$ norm for fusion
frames.

A signal sparse in a fusion frame can be sampled using very few random
projections and exactly reconstructed using a convex optimization that
minimizes this mixed $\ell_1/\ell_2$ norm. The provided sampling
conditions generalize coherence and RIP conditions used in standard CS
theory. It is demonstrated that they are sufficient to guarantee
sparse recovery of any signal sparse in our model. Moreover, a
probabilistic analysis is provided using a stochastic model on the
sparse signal that shows that under very mild conditions the
probability of recovery failure decays exponentially with increasing
dimension of the subspaces.
\end{abstract}


\IEEEpeerreviewmaketitle

\section{Introduction}
\IEEEPARstart{C}{ompressed} Sensing (CS) has recently emerged as a
very powerful field in signal processing, enabling the acquisition
of signals at rates much lower than previously thought
possible~\cite{CRT06b,Don06}. To achieve such performance, CS
exploits the structure inherent in many naturally occurring and
man-made signals. Specifically, CS uses classical signal
representations and imposes a sparsity model on the signal of
interest. The sparsity model, combined with randomized linear
acquisition, guarantees that non-linear reconstruction can be used
to efficiently and accurately recover the signal.

Fusion frames are recently emerged mathematical structures that can
better capture the richness of natural and man-made signals compared
to classically used representations~\cite{CKL08}. In particular,
fusion frames generalize frame theory by using subspaces in the place
of vectors as signal building blocks. Thus signals can be represented
as linear combinations of components that lie in particular, and often
overlapping, signal subspaces. Such a representation provides
significant flexibility in representing signals of interest compared
to classical frame representations.

In this paper we extend the concepts and methods of Compressed Sensing
to fusion frames. In doing so we demonstrate that it is possible to
recover signals from underdetermined measurements if the signals lie
only in very few subspaces of the fusion frame. Our generalized model
does not require that the signals are sparse within each subspace. The
rich structure of the fusion frame framework allows us to characterize
more complicated signal models than the standard sparse or
compressible signals used in compressed sensing techniques. This paper
complements and extends our work in~\cite{BKRSPIE09}.

Introducing sparsity in the rich fusion frame model and introducing
fusion frame models to the Compressed Sensing literature is a major
contribution of this paper. We extend the results of the standard
worst-case analysis frameworks in Compressed Sensing, using the
sampling matrix {\em null space property (NSP)}, {\em coherence}, and
{\em restricted isometry property (RIP)}. In doing so, we extend the
definitions to fusion NSP, fusion coherence and fusion RIP to take
into account the differences of the fusion frame model. The three
approaches provide complementary intuition on the differences between
standard sparsity and block-sparsity models and sparsity in fusion
frame models. We note that in the special case that the subspaces in our model
all have the same dimension, most of our results follow from previous
analysis of block-sparsity models \cite{elmi09,EKB10}. 
But in the general case of different dimensions they are new.

Our understanding of the problem is further enhanced by the
probabilistic analysis. As we move from standard sparsity to fusion
frame or other vector-based sparsity models, worst case analysis
becomes increasingly pessimistic. The probabilistic analysis provides
a framework to discern which assumptions of the worst case model
become irrelevant and which are critical. It further demonstrates the
significance of the angles between the subspaces comprising the fusion
frame. Although our analysis is inspired by the model and the analysis
in~\cite{elra09}, the tools used in that work do not extend to the
fusion frame model. The analysis presented in
Sec.~\ref{sec:average_case} is the second major contribution of our
paper.

In the remainder of this section we provide the motivation behind our
work and describe some possible
applications. Section~\ref{sec:background} provides some background on
Compressed Sensing and on fusion frames to serve as a quick reference
for the fundamental concepts and our basic notation. In
Section~\ref{sec:formulation} we formulate the problem, establish the
additional notation and definitions necessary in our development, and
state the main results of our paper.  We further explore the
connections with existing research in the field, as well as possible
extensions. In Section~\ref{sec:coherence} we prove deterministic
recovery guarantees using the properties of the sampling
matrix. Section~\ref{sec:average_case} presents the probabilistic
analysis of our model, which is more appropriate for typical usage
scenarios. We conclude with a discussion of our results.

\subsection{Motivation}
As technology progresses, signals and computational sensing
equipment becomes increasingly multidimensional. Sensors are being
replaced by sensor arrays and samples are being replaced by
multidimensional measurements. Yet, modern signal acquisition theory
has not fully embraced the new computational sensing paradigm.
Multidimensional measurements are often treated as collections of
one-dimensional ones due to the mathematical simplicity of such
treatment. This approach ignores the potential information and
structure embedded in multidimensional signal and measurement
models.

Our ultimate motivation is to provide a better understanding of more
general mathematical objects, such as vector-valued data
points~\cite{BDE07}. Generalizing the notion of sparsity is part of
such understanding. Towards that goal, we demonstrate that the
generalization we present in this paper encompasses joint sparsity
models~\cite{fora08,gisttr06-1} as a special case. Furthermore, it is
itself a special case of block-sparsity
models~\cite{Peotta07,Kowalski:2009fk,elmi09,EKB10}, with significant
additional structure. 

\subsection{Applications}
Although the development in this paper provides a general
theoretical perspective, the principles and the methods we develop
are widely applicable. In particular, the special case of joint (or
simultaneous) sparsity has already been widely used in
radar~\cite{MZ06}, sensor arrays~\cite{Malioutov03}, and MRI pulse
design~\cite{Zelinski08}. In these applications a mixed
$\ell_1/\ell_2$ norm was used heuristically as a sparsity proxy.
Part of our goals in this paper is to provide a solid theoretical
understanding of such methods.

In addition, the richness of fusion frames allows the application of
this work to other cases, such as target recognition and music
segmentation. The goal in such applications is to identify, measure,
and track targets that are not well described by a single vector but
by a whole subspace. In music segmentation, for example, each note
is not characterized by a single frequency, but by the subspace
spanned by the fundamental frequency of the instrument and its
harmonics~\cite{daud06}. Furthermore, depending on the type of
instrument in use, certain harmonics might or might not be present
in the subspace. Similarly, in vehicle tracking and identification,
the subspace of a vehicle's acoustic signature depends on the type
of vehicle, its engine and its tires~\cite{ccm09}. Note that in both
applications, there might be some overlap in the subspaces which
distinct instruments or vehicles occupy.

Fusion frames are quite suitable for such representations. The
subspaces defined by each note and each instrument or each tracked
vehicle generate a fusion frame for the whole space. Thus the fusion
frame serves as a dictionary of targets to be acquired, tracked, and
identified. The fusion frame structure further enables the use of
sensor arrays to perform joint source identification and localization
using far fewer measurements than a classical sampling framework. In
Sec.~\ref{sec:previous} we provide a stylized example that
demonstrates this potential.

We also envision fusion frames to play a key role in video
acquisition, reconstruction and compression applications such
as~\cite{VRR10}. Nearby pixels in a video exhibit similar sparsity
structure locally, but not globally. A joint sparsity model such
as~\cite{fora08,gisttr06-1} is very constraining in such
cases. On the other hand, subspace-based models for different parts of
an image significantly improve the modeling ability compared to the
standard compressed sensing model.

\subsection{Notation}

Throughout this paper
$\|\x\|_p=\left(\sum_ix_i^p\right)^{1/p},~p>0$ denotes the standard
$\ell_p$ norm. The operator norm of a matrix $A$ from $\ell_p$ into $\ell_p$
is written as $\|A\|_{p \to p} = \max_{\|x\|_p\leq 1} \|Ax\|_p$.

\section{Background}
\label{sec:background}
\subsection{Compressed Sensing}
Compressed Sensing (CS) is a recently emerged field in signal
processing that enables signal acquisition using very few
measurements compared to the signal dimension, as long as the signal
is sparse in some basis. It predicts that a signal $\x\in \RR^N$
with only $k$ non-zero coefficients can be recovered from only
$n={\cal{O}}(k\log(N/k))$ suitably chosen linear non-adaptive
measurements, compactly represented by
\[
\y=\A\x, \quad \y\in \RR^n, \A\in\RR^{n\times N}.
\]
A necessary condition for exact signal recovery of all $k$-sparse $\x$ is that
\[
\A\z\ne 0 \quad \mbox{~for~all~} \z\ne 0,\|\z\|_0\le 2k,
\]
where the $\ell_0$ `norm,' $\|\x\|_0$, counts the number of non-zero
coefficients in $\x$. In this case, recovery is possible using the following
combinatorial optimization,
\[
\widehat{\x}=\argmin_{\x\in\RR^N}\|\x\|_0\mbox{~subject~to~}\y=\A\x.
\]
Unfortunately this is an NP-hard problem \cite{na95} in general,
hence is infeasible. 

Exact signal recovery using computationally tractable methods can be
guaranteed if the measurement matrix $\A$ satisfies the null space
property (NSP)~\cite{cdd09,Z05}, i.e., if for all support sets $S
\subset \{1,\hdots,N\}$ of cardinality at most $k$,
\[
\|\h_S\|_1 < \frac{1}{2} \|\h\|_1 \quad \mbox{ for all } \h
\in \mathrm{nullsp}(\A) \setminus \{0\},
\]
where $\h_S$ denotes the vector which coincides with $\h$ on the index
set $S$ and is zero outside $S$.

If the coherence of $\A$ is sufficiently small, the measurement matrix
satisfies the NSP and, therefore, exact recovery is
guaranteed~\cite{tr04,BDE07}. The {\em coherence} of a matrix $\A$
with unit norm columns $\a_i$, $\|\a_i\|_2 = 1$, is defined as
\begin{equation}
  \mu=\max_{i\ne j}\left|\langle\a_i,\a_j\rangle\right|.
  \label{eq:dict_coherence}
\end{equation}
Exact signal recovery is also guaranteed if
$\A$ obeys a {\em restricted isometry property (RIP) of order $2k$} \cite{CRT06b},
i.e., if there exists a constant $\delta_{2k} \in (0,1)$ such that for all
$2k$-sparse signals $\x$
\[
(1-\delta_{2k})\|\x\|_2^2\le\|\A\x\|_2^2\le(1+\delta_{2k})\|\x\|_2^2.
\]
We note the relation $\delta_{2k} \leq (k-1) \mu$, which easily
follows from Gershgorin's theorem. 
If any of the above properties hold, then the following convex
optimization program exactly recovers the signal from the measurement
vector $\y$,
\[
\widehat{\x}=\argmin_{\x\in\RR^N}\|\x\|_1\mbox{~subject~to~}\y=\A\x.
\]
A surprising result is that random matrices with sufficient number
of rows can achieve small coherence and small RIP constants with
overwhelmingly high probability.

A large body of literature extends these results to measurements of
signals in the presence of noise, to signals that are not exactly
sparse but compressible \cite{CRT06b}, to several types of
measurement matrices \cite{cata06,ra08-1,pfra07,ra09,badularotr10}
and to measurement models beyond simple sparsity \cite{baceduhe08}.

\subsection{Fusion Frames}
\label{sec:FFBackground}
Fusion frames are generalizations of frames that provide a richer
description of signal spaces. A {\em fusion frame} for $\RR^M$ is a
collection of subspaces $\mathcal{W}_j\subseteq\RR^M$ and
associated weights $v_j$, compactly denoted by
$(\mathcal{W}_j,v_j)_{j=1}^{N}$, that satisfies
\[
A\|\x\|_2^2\le\sum_{j=1}^Nv_j^2\left\|\P_j \x \right\|_2^2\le
B\|\x\|_2^2
\]
for some universal fusion frame bounds $0<A\le B<\infty$ and for all
$\x\in\RR^M$, where $\P_j$ denotes the orthogonal projection onto
the subspace $\mathcal{W}_j$. We use $m_j$ to denote the dimension of
the $j$th subspace $\mathcal{W}_j$, $j=1,\ldots, N$. A frame is a
special case of a fusion frame in which all the subspaces
$\mathcal{W}_j$ are one-dimensional (i.e., $m_j=1,~j=1,\ldots,N$), and
the weights $v_j$ are the norms of the frame vectors. For finite $M$
and $N$, the definition reduces to the requirement that the subspace
sum of $\mathcal{W}_j$ is equal to $\RR^M$.

The generalization to fusion frames allows us to capture interactions
between frame vectors to form specific subspaces that are not possible
in classical frame theory. Similar to classical frame theory, we call
the fusion frame {\em tight} if the frame bounds are equal, $A=B$. If
the fusion frame has $v_j=1,~j=1,\ldots,N$, we call it a {\em
  unit-norm} fusion frame.  In this paper, we will in fact restrict to
the situation of unit-norm fusion frames, since the anticipated
applications are only concerned with membership in the subspaces and
do not necessitate a particular weighting.

Dependent on a fusion frame $(\mathcal{W}_j,v_j)_{j=1}^{N}$ we define the Hilbert space $\cH$ as
\begin{eqnarray*}
\cH &=& \{(\x_j)_{j=1}^N : \x_j \in \cW_j \mbox{ for all }j=1,\ldots,N\}\notag\\
&\subseteq& \RR^{M \times N} (\mathrm{or}~\RR^{MN}).\notag
\end{eqnarray*}
We should point out that depending on the use, $\mathbf{x}$ can be
represented as a very long vector or as a matrix. However, both
representations are just rearrangements of vectors in the same Hilbert
space, and we use them interchangeably in the manuscript.

Finally, let $\U_j\in \RR^{M\times m_j}$ be a known but
otherwise arbitrary matrix, the columns of which form an orthonormal
basis for $\mathcal{W}_j$, $j=1,\ldots, N$, that is
$\U_j^T\U_j=\I_{m_j}$, where $\I_{m_j}$ is the $m_j\times m_j$
identity matrix, and $\U_j\U_j^T=\P_j$.

The {\em fusion frame mixed $\ell_{q,p}$ norm} is defined as
\begin{equation}
 \left\|(\x_j)_{j=1}^N\right\|_{q,p}\equiv
 \left(\sum_{j=1}^N\left(v_j\|\x_j\|_q\right)^p\right)^{1/p},
\end{equation}
where $(v_j)_{j=1}^N$ are the fusion frame weights.
Furthermore, for a sequence $\c = (\c_j)_{j=1}^N$, $\c_j \in
\RR^{m_j}$, we similarly define the mixed norm
\[
\|\c\|_{2,1} = \sum_{j=1}^N \|\c_j\|_2.
\]
The $\ell_{q,0}$--`norm' (which is actually not even a quasi-norm) is
defined as
\[
\|\x\|_{q,0} = \# \{j : \x_j \neq 0\},
\]
independent of $q$. For the remainder of the paper we use $q=2$ just
for the purpose of notation and to distinguish $\ell_{2,0}$ from the
$\ell_0$ vector 'norm'. We call a vector $\x \in \cH$ {\em
  $k$-sparse}, if $\|\x\|_{2,0} \le k$.

\section{Sparse Recovery of Fusion Frame Vectors}
\label{sec:formulation}
\subsection{Measurement Model}
We now consider the following scenario. Let $\x^0=(\x_j^0)_{j=1}^N \in
\cH$, and assume that we only observe $n$ linear combinations of those
vectors, i.e., there exist some scalars $a_{ij}$ satisfying 
$\norm{(a_{ij})_{i=1}^n}_2 = 1$ for all $j=1,\ldots,N$ such that we
observe \beq \label{eq:linearcomb} \y = (\y_i)_{i=1}^n =
\left(\sum_{j=1}^N a_{ij} \x_j^0\right)_{i=1}^n \in \cK, \eeq where
$\cK$ denotes the Hilbert space
\[
\cK = \{(\y_i)_{i=1}^n : \y_i \in \RR^M \mbox{ for all
}i=1,\ldots,n\}.
\]

We first notice that \eqref{eq:linearcomb} can be rewritten as
\[
\y = \A_\I \x^0, \quad \mbox{where } \A_\I = (a_{ij} \I_M)_{1 \le i
\le n,\, 1 \le j \le N},
\]
i.e., $\A_\I$ is the matrix consisting of the blocks $a_{ij} \I_M$.

\subsection{Reconstruction using Convex Optimization}
We now wish to recover $\x^0$ from the measurements $\y$. If we impose
conditions on the sparsity of $\x^0$, it is suggestive to consider the
following minimization problem,
\begin{eqnarray*}
\hat{\x} &= & \argmin_{\x \in \cH} \norm{\x}_{2,0} \\ 
&& \mbox{subject to }
\sum_{j=1}^N a_{ij}\x_j = \y_i \mbox{ for all } i=1,\ldots,n.
\end{eqnarray*}
Using the matrix $\A_\I$, we can rewrite this optimization problem
as
\[
{\sc (P_0)}\quad\hat{\x}= \argmin_{\x \in \cH} \norm{\x}_{2,0} \quad \mbox{
subject to } \A_\I\x = \y.
\]
However, this problem is NP-hard~\cite{na95} and, as proposed in
numerous publications initiated by~\cite{CDS01}, we prefer to employ
$\ell_1$ minimization techniques.  This leads to the investigation
of the following minimization problem,
\[
\hat{\x}= \argmin_{\x \in \cH} \norm{\x}_{2,1}\quad  \mbox{ subject to }
\A_\I\x = \y.
\]
Since we minimize over all $\x = (\x_j)_{j=1}^N \in \cH$ and
certainly $\P_j \x_j = \x_j$ by definition, 
we can rewrite this minimization problem as
\[
{\sc (\tilde{P}_1)}\quad\hat{\x}= \argmin_{\x \in \cH} \norm{\x}_{2,1}
\quad \mbox{ subject to } \A_\P\x = \y,
\]
where \beq \label{eq:definitionAP} \A_\P = (a_{ij} \P_j)_{1 \le i
\le n,\, 1 \le j \le N}. \eeq

Problem ${\sc (\tilde{P}_1)}$ bears difficulties to implement since
minimization runs over $\cH$. Still, it is easy to see that ${\sc
  (\tilde{P}_1)}$ is equivalent to the optimization problem
\begin{eqnarray} 
{\sc (P_1)}\quad(\hat{\c}_j)_j & = & \argmin_{\c_j \in
\RR^{m_j}} \norm{(\U_j\c_j)_{j=1}^N}_{2,1}\label{eq:optim_ff} \\ 
&&\mbox{subject to }
\A_\I(\U_j\c_j)_j = \y, \notag
\end{eqnarray} 
where then
$\hat{\x} = (\U_j\hat{\c}_j)_{j=1}^N$. This particular form ensures
that the minimizer lies in the collection of subspaces $(\cW_j)_{j=1}^N$
while minimization is performed over $\c_j \in \RR^{m_j}$ for all $
j=1, \ldots, N$ and $\sum_jm_j\le MN$, hence feasible.

Finally, by rearranging \eqref{eq:optim_ff}, the optimization problems, invoking
the $\ell_0$-`norm' and $\ell_1$-norm, can be rewritten using matrix-only notation
as
\[
\mbox{{\sc ($P_0$)}} \quad \hat{\c} = \argmin_{\c} \norm{\c}_{2,0}
\mbox{ subject to } \Y = \A \U(\c)
\]
and
\[
\mbox{{\sc ($P_1$)}} \quad \hat{\c} = \argmin_{\c} \norm{\c}_{2,1}
\mbox{ subject to } \Y = \A \U(\c),
\]
in which
\begin{eqnarray*}
  \U(\c) & = & \left( \begin{array}{c}
    \c_1^T \U_1^T\\
    \hline\vspace*{-0.5cm} \\
    \vdots\\
    \hline\vspace*{-0.35cm}\\
    \c_N^T \U_N^T
  \end{array} \right) \in \RR^{N \times M},~
  \Y = \left( \begin{array}{c}
    \y_1\\
    \hline \vspace*{-0.5cm}\\
    \vdots\\
    \hline\vspace*{-0.5cm}\\
    \y_n
  \end{array} \right) \in \RR^{n \times M},\\ 
  \A& = & (a_{ij}) \in
  \RR^{n \times N}, \quad \c_j \in \RR^{m_j},
  \; ~\mbox{and~}\y_i \in \RR^{M}.
\end{eqnarray*}
Hereby, we additionally used that $\|\U_j \c_j\|_2 = \|\c_j\|_2$ by
orthonormality of the columns of $\U_j$. We follow this notation for the
remainder of the paper.

\subsection{Worst Case Recovery Conditions}
To guarantee that~${\sc (\tilde{P}_1)}$ always recovers the original
signal $\x^0$, we provide three alternative conditions, in line with
the standard CS literature~\cite{BDE07,cata06,CRT06b,cdd09,DET06,grni03,ra09-1,tr04}. 
Specifically, we define the fusion
null space property, the fusion coherence, and the fusion restricted
isometry propery (FRIP). These are fusion frame versions of the
null space property, the coherence and the RIP, respectively.

\begin{definition} The pair $(\A,
(\cW_j)_{j=1}^N)$, with a matrix $\A \in \RR^{n \times N}$ and a fusion
frame $(\cW_j)_{j=1}^N$ is said to satisfy the {\em fusion null space
  property} if
\[
\|\h_S\|_{2,1} < \frac{1}{2} \|\h\|_{2,1} \quad \mbox{ for all } \h \in {\cal{N}} \setminus \{0\},
\]
for all support sets $S \subset \{1,\hdots,N\}$ of cardinality at most
$k$. Here ${\cal{N}}$ denotes the null space $\{\h = (\h_j)_{j=1}^N:
\h_j \in \RR^{m_j}, \A \U(\h) = 0\}$, and $\h_S$ denotes the vector
which coincides with $\h$ on the index set $S$ and is zero outside
$S$.
\end{definition}

\begin{definition}
\label{def:fusion_coherence} The {\em fusion coherence} of a matrix
$\A \in \RR^{n \times N}$ with normalized `columns' $(\a_j =
\a_{\cdot , j})_{j=1}^N$ and a fusion frame $(\cW_j)_{j=1}^N$
for $\RR^{M}$ is given by
\[
\mu_f=\mu_f(\A,(\cW_j)_{j=1}^N) = \max_{j \neq k} \left[
\absip{\a_j}{\a_k} \cdot \norm{\P_j\P_k}_{2 \to 2}\right],
\]
where $\P_j$ denotes the orthogonal projection onto $\cW_j$, $j=1,\ldots,N$.
\end{definition}

\begin{definition}
\label{def:frip}
Let $\A \in \RR^{n \times N}$ and $(\cW_j)_{j=1}^N$
  be a fusion frame for $\RR^M$ and $\A_\P$ as defined in
  \eqref{eq:definitionAP}. The {\em fusion restricted isometry
    constant} $\delta_k$ is the smallest constant such that
\[
(1-\delta_k) \|\z\|_{2,2}^2 \leq \|\A_\P \z\|_{2,2}^2 \leq (1+\delta_k) \|\z\|_{2,2}^2
\]
for all $\z = (\z_1, \z_2, \hdots, \z_N) \in \RR^{MN}$, $\z_j \in
\RR^M$, of sparsity $\|\z\|_0\leq k$.
\end{definition}

In the case that the subspaces have the same dimension the definitions
of fusion coherence and fusion RIP coincide with those of block
coherence and block RIP introduced in \cite{elmi09,EKB10}. This is
expected, as we discuss in~Sec.~\ref{sec:previous}, since the fusion
sparsity model is a special case of general block sparsity models.

Using those definitions, we provide three alternative recovery
conditions, also in line with standard CS results. We first state the
characterization via the fusion null space property.

\begin{theorem}
\label{lem:FNSP}
Let $\A \in \RR^{n \times N}$ and $(\cW_j)_{j=1}^N$ be a fusion frame.
Then all $\c = (\c_j)_{j=1}^N$, $\c_j \in \RR^{m_j}$, with $\|\c\|_{2,0}
\leq k$ are the unique solution to $(P_1)$ with $\Y = \A \U(\c)$ if
and only if $(\A, (\cW_j)_{j=1}^N)$ satisfies the fusion null space
property of order $k$.
\end{theorem}

While the fusion null space property characterizes recovery, it is somewhat difficult to check in practice. The fusion coherence
is much more accessible for a direct calculation, and the next result states a corresponding sufficient condition.
In the case, that the subspaces have the same
dimension the theorem below reduces to Theorem 3 in \cite{EKB10} on block-sparse recovery.

\begin{theorem}
\label{theo:main}
Let $\A \in \RR^{n \times N}$ with normalized columns
$(\a_j)_{j=1}^N$, let $(\cW_j)_{j=1}^N$ be a fusion frame in
$\RR^{M}$, and let $\Y \in \RR^{n \times M}$. If there exists a
solution $\c^0$ of the system $\A\U(\c)=\Y$ satisfying
\beq \label{eq:sparsitycondition} \norm{\c^0}_{2,0} < \frac12 (1 +
\mu_f^{-1}), \eeq then this solution is the unique solution of $(P_0)$
as well as of $(P_1)$.
\end{theorem}

Finally, we state a sufficient condition based on the fusion RIP, which allows stronger recovery results, but is more difficult to evaluate than the
fusion coherence. The constant $1/3$ below is not optimal and can certainly be improved,
but our aim was rather to have a short proof. Furthermore, in the case that all the subspaces have the same dimension, the fusion frame
setup is equivalent to the block-sparse case, for which the next theorem is implied by Theorem 1 in \cite{elmi09}.

\begin{theorem}
\label{theo_FRIP1}
Let $(\A,(\cW_j)_{j=1}^N)$ with fusion frame restricted isometry
constant $\delta_{2k} < 1/3$. Then $(P_1)$ recovers all $k$-sparse
$\c$ from $\Y = \A \U (\c)$.
\end{theorem}

\subsection{Probability of Correct Recovery}
\label{sec:prob_result_intro}
Intuitively, it seems that the higher the dimension of the subspaces $\cW_j$, the `easier' the
recovery via the $\ell_1$ minimization problem \cite{elra09} should
be. However, it turns out that this intuition only holds true
if we consider a probabilistic analysis. 
Thus we provide a probabilistic signal model and a
typical case analysis for the special case where all the subspaces
have the same dimension $m=m_j$ for all $j$. Inspired by the probability model in \cite{elra09}, we will
assume that on the $k$-element support set $S=\supp(\U(\c)) = \{j_1,
\ldots, j_{k}\}$ the entries of the vectors $\c_j$, $j \in S$, are independent
and follow a normal distribution,
\beq
\label{eq:probmodel} \U(\c)_S
= \left( \begin{array}{c}
\X_1^T \U_{j_1}^T\\
\hline\vspace*{-0.5cm} \\
\vdots\\
 \hline\vspace*{-0.45cm}\\
\X_{k}^T \U_{j_{k}}^T
\end{array} \right) \in \RR^{k \times M},
\eeq where $\X = (\X_1^T \ldots \X_{k}^T)^T \in \RR^{Nm}$ is a
Gaussian random vector, i.e., all entries are independent standard
normal random variables.

Our probabilistic result shows that the failure probability for
recovering $\U(\c)$ decays exponentially fast with growing dimension
$m$ of the subspaces. Interestingly, the quantity $\theta$ involved in
the estimate is again dependent on the `angles' between subspaces and
is of the flavor of the fusion coherence in
Def.~\ref{def:fusion_coherence}. Since the block-sparsity model can be seen as a special
case of the fusion frame sparsity model, the theorem clearly applies also to this scenario.

\begin{theorem}\label{thm:main}
Let $S \subseteq \{1,\ldots,N\}$ be a set of cardinality $k$ and
suppose that $\A \in \RR^{n \times N}$ satisfies
\beq
\label{eq:main_eq1} \norm{\A_S^\dagger \A_{\cdot,j}}_2 \le \alpha <
1 \quad \mbox{for all } j \not\in S.
\eeq
Let $(\cW_j)_{j=1}^N$ be a fusion frame with associated orthogonal
bases $(\U_j)_{j=1}^N$ and orthogonal projections $(\P_j)_{j=1}^N$,
and let $\Y \in \RR^{n \times M}$. Further, let $\c_j \in \RR^{m_j}$,
$j = 1 \ldots, N$ with $S = \supp(\U(\c))$ such that the
coefficients on $S$ are given by \eqref{eq:probmodel}, and let
$\theta$ be defined by
\[
\theta = 1+\max_{i \in S} \sum_{j \in S, j \neq i} \lambda_{\max{}}(\P_i \P_j)^{1/2}.
\]
Choose $\delta \in (0,1-\alpha^2)$. Then with probability at least
\[
1-(N-k) \exp\left(-\frac{(\sqrt{1-\delta}
-\alpha)^2}{2\alpha^2\theta} m\right) - k
\exp\left(-\frac{\delta^2}{4} m\right)
\]
the minimization problem $(P_1)$
recovers $\U(\c)$ from $\Y = \A \U(\c)$. In particular, the failure
probability can be estimated by
\[
N \,\exp\left(-\left(\max_{\delta \in (0,1-\alpha^2)} \min\left\{
\frac{(\sqrt{1-\delta} -\alpha)^2}{2\alpha^2\theta},
\frac{\delta^2}{4}\right\}\right) m\right).
\]
\end{theorem}
Let us note that \cite[Section V]{elra09} provides several mild
conditions that imply \eqref{eq:main_eq1}. We exemplify one of these.
Suppose that the columns of $A \in \RR^{n \times N}$ form a unit norm
tight frame with (ordinary) coherence $\mu \leq c / \sqrt{n}$. (This
condition is satisfied for a number of explicitly given matrices, see
also \cite{elra09}.) Suppose further that the support set $S$ is
chosen uniformly at random among all subsets of cardinality $k$. Then
with high probability \eqref{eq:main_eq1} is satisfied provided $k
\leq C_\alpha n$, see Theorem 5.4 and Section V.A in
\cite{elra09}. This is in sharp contrast to deterministic recovery
guarantees based on coherence, which cannot lead to better bounds than
$k \leq C \sqrt{n}$. Further note, that the quantity $\theta$ is
bounded by $k$ in the worst case, but maybe significantly smaller if
the subspaces are close to orthogonal. Clearly, $m$ can only be varied
by changing the fusion frame model, and $\theta$ may also change in
this case.  While, in principle, $\theta$ may grow with $m$, typically
$\theta$ actually decreases; for instance if the subspaces are chosen
at random.  In any case, since always $\theta \leq k$, the failure
probability of recovery decays exponentially in the dimension $m$ of
the subspaces.

\subsection{Relation with Previous Work}
\label{sec:previous}

A special case of the problem above appears when all subspaces
$(\mathcal{W}_j)_{j=1}^{N}$ are equal and also equal to the ambient
space $\cW_j=\RR^M$ for all $j$. Thus, $\P_j=\I_M$ and the
observation setup of Eq.~\eqref{eq:linearcomb} is identical to the
matrix product
\begin{equation} \Y=\A\X,
\quad \mbox{where }
 \X = \left( \begin{array}{c}
    \x_1\\
    \hline \vspace*{-0.5cm}\\
    \vdots\\
    \hline\vspace*{-0.5cm}\\
    \x_N
  \end{array} \right) \in \RR^{N \times M}.
 \label{eq:joint_block}
\end{equation}
This special case is the same as the well studied joint sparsity
setup of \cite{bb05,fora08,gisttr06-1,grrascva08,elra09} in which a
collection of $M$ sparse vectors in $\RR^N$ is observed through
the same measurement matrix $\A$, and the recovery assumes that all
the vectors have the same sparsity structure. The use of mixed
$\ell_1/\ell_2$ optimization has been proposed and widely used in
this case.

The following simple example illustrates how a fusion frame model can
reduce the sampling required compared to simple joint sparsity
models. Consider the measurement scenario of \eqref{eq:joint_block},
rewritten in an expanded form:
\begin{eqnarray*}
  \left[
    \begin{array}{c}
      -\y_1^T-\\
      \vdots\\
      -\y_n^T-\\
    \end{array}
    \right]=
  \left[
    \begin{array}{cccc}
      |&|&&|\\
      \a_1&\a_2&\ldots&\a_N\\
      |&|&&|
    \end{array}
    \right]
  \left[    \begin{array}{c}
      -\x_1^T-\\
      -\x_2^T-\\
      \vdots\\
      -\x_N^T-\\
    \end{array}
\right],
\end{eqnarray*}
where $\x_i,\y_i\in\RR^M$ and $\x_i\in\cW_i$. Using the fusion
frames notation, $\X$ is a fusion frame representation of our acquired
signal, which we assume sparse as we defined in
Sec.~\ref{sec:FFBackground}. For the purposes of the example, suppose
that only $\x_1$ and $\x_2$ have significant content and the remaining
components are zero. Thus we only focus on the measurement vectors
$\a_1$ and $\a_2$ and their relationship, with respect to the
subspaces $\cW_1$ and $\cW_2$ where $\x_1$ and $\x_2$ lie in.

Using the usual joint sparsity models, would require that $\a_1$ and
$\a_2$ have low coherence \eqref{eq:dict_coherence}, even if prior
information about $\cW_1$ and $\cW_2$ provided a better signal
model. For example, if we know from the problem formulation that the
two components $\x_1$ and $\x_2$ lie in orthogonal subspaces
$\cW_1\perp\cW_2$, we can select $\a_1$ and $\a_2$ to be identical,
and still be able to recover the signal. If, instead, $\cW_1$ and
$\cW_2$ only have a common overlapping subspace we need $\a_1$ and
$\a_2$ to be incoherent only when projected on that subspace, as
measured by fusion coherence, defined in
Def.~\ref{def:fusion_coherence}, and irrespective of the
dimensionality of the two subspaces and the dimensionality of their
common subspace. This is a case not considered in the existing
literature.

The practical applications are significant. For example, consider a
wideband array signal acquisition system in which each of the signal
subspaces are particular targets of interest from particular
directions of interest (e.g. as an extremely simplified stylized
example consider $\cW_1$ as the subspace of friendly targets and
$\cW_2$ as the subspace of enemy targets from the same direction). If
two kinds of targets occupy two different subspaces in the signal
space, we can exploit this in the acquisition system. This opens the
road to subspace-based detection and subspace-based target
classification dictionaries.

Our formulation is a special case of the block sparsity
problem~\cite{Peotta07,Kowalski:2009fk,EKB10}, where we impose a
particular structure on the measurement matrix $\A$. This relationship
is already known for the joint sparsity model, which is also a special
case of block sparsity. In other words, the fusion frame formulation
we examine here specializes block sparsity problems and generalizes
joint sparsity ones. 

For the special case of fusion frames in which all the subspaces
$\mathcal{W}_j$ have the same dimension, our definition of coherence
can be shown to be essentially the same (within a constant scaling
factor) as the one in~\cite{EKB10} when the problem is reformulated as
a block sparsity one. Similarly, for the same special case our
definition of the NSP becomes similar to the one in~\cite{SPH09}.

\begin{figure}[t]
   \centerline{\includegraphics[width=\linewidth]{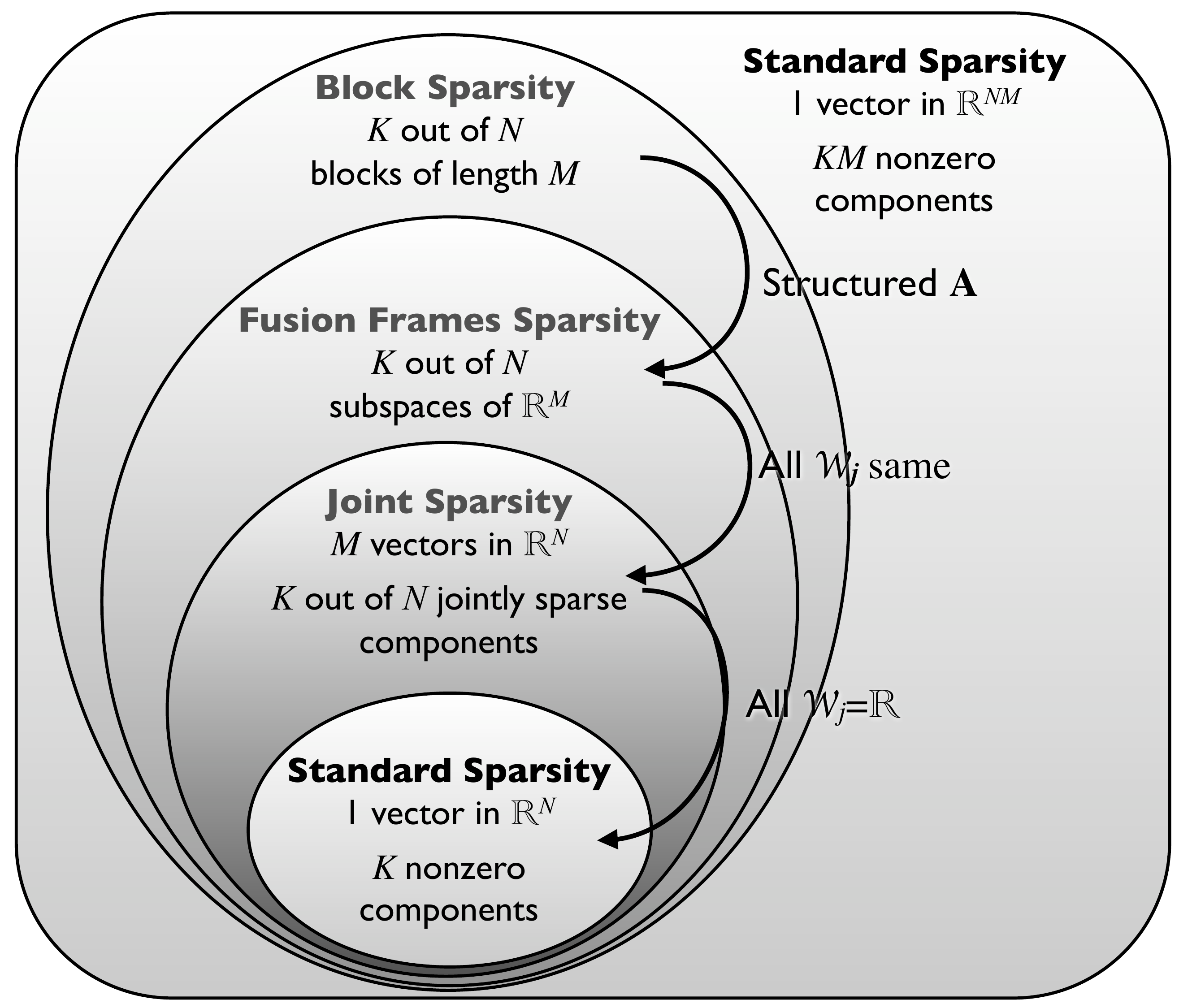}}
  \caption{Hierarchy of Sparsity Models}
  \label{fig:model_venn}
\end{figure}

We would also like to note that the hierarchy of such sparsity
problems depends on their dimension. For example, a joint sparsity
problem with $M=1$ becomes the standard sparsity model. In that
sense, joint sparsity models generalize standard sparsity models.
The hierarchy of sparsity models is illustrated in the Venn diagram
of Fig.~\ref{fig:model_venn}.

\subsection{Extensions}
Several extensions of this formulation and the work in this paper
are possible, but beyond our scope.
For example, the
analysis we provide is in the exactly sparse, noiseless case. As
with classical compressed sensing, it is possible to accommodate
sampling in the presence of noise. It is also natural to consider
the extension of this work to sampling signals that are not
$k$-sparse in a fusion frame representation but can be very well
approximated by such a representation. (However, see Section \ref{RIP_Remarks}.)

The richness of fusion frames also allows us to consider richer
sampling matrices. Specifically, it is possible to consider sampling
operators consisting of different matrices, each operating on a
separate subspace of the fusion frame. Such extensions open the use of
$\ell_1$ methods to general vector-valued mathematical objects, to the
general problem of sampling such objects \cite{BDE07}, and to general
model-based CS problems \cite{baceduhe08}.

\section{Deterministic Recovery Conditions}
\label{sec:coherence}

In this section we derive conditions on $\c^0$ and $\A$ so that $\c^0$
is the unique soÜlution of ($P_0$) as well as of ($P_1$). Our approach
uses the generalized notions of {\em null space property}, {\em
  coherence} and the {\em restricted isometry property}, all commonly
used measures of morphological difference between the vectors of a
measuring matrix. 

\subsection{Fusion Null Space Property}
We first prove Thm.~\ref{lem:FNSP}, which demonstrates that the fusion
NSP guarantees recovery, similarly to the standard CS
setup~\cite{cdd09}. This notion will also be useful later to prove
recovery bounds using the fusion coherence and using the fusion
restricted isometry constants.

\begin{IEEEproof}[Proof of Theorem \ref{lem:FNSP}] Assume first that the fusion NSP holds.
Let $\c^0$ be a vector with $\|\c^0\|_{2,0} \leq k$, and let $\c^1$ be an
arbitrary solution of the system $\A\U(\c)=\Y$, and set
\[
\h = \c^1-\c^0.
\]
Letting $S$ denote the support of $\c^0$, we obtain
\begin{eqnarray*}
\norm{\c^1}_{2,1} - \norm{\c^0}_{2,1} 
&= & \norm{\c^1_{S}}_{2,1} +
\norm{\c^1_{S^c}}_{2,1} - \norm{\c^0_S}_{2,1}\\
& \ge &
\norm{\h_{S^c}}_{2,1} - \norm{\h_S}_{2,1}.
\end{eqnarray*}
This term is greater than zero for any $\h \neq
0$ provided that
\begin{equation} \label{eq:nullspace}
\norm{\h_{S^c}}_{2,1} > \norm{\h_S}_{2,1}
\end{equation}
or, in other words,
\beq \label{eq:inequality_h} \tfrac12
\norm{\h}_{2,1} > \norm{\h_S}_{2,1},
\eeq
which is ensured by the fusion NSP.

Conversely, assume that all vectors $\c$ with $\|\c\|_0 \leq k$ are recovered
using $(P_1)$. Then, for any $\h \in {\cal{N}} \setminus \{0\}$ and any
$S \subset \{1,\hdots,N\}$ with $|S| \leq k$, the $k$-sparse vector $\h_S$ is
the unique minimizer of $\|\c\|_{2,1}$ subject to $\A \c = \A \h_S$. Further, observe that
$\A (-\h_{S^c}) = \A (\h_S)$ and $-\h_{S^c} \neq \h_S$, since $\h \in {\cal{N}} \setminus \{0\}$.
Therefore, $\|\h_S\|_{2,1} <\|\h_{S^c}\|_{2,1}$, which is equivalent to the fusion
NSP because $\h$ was arbitrary.
\end{IEEEproof}

\subsection{Fusion Coherence}

The fusion coherence is an adaptation of the coherence notion to our
more complicated situation involving the {\em angles} between the
subspaces generated by the bases $\U_j$, $j=1,\ldots,N$. In other
words, here we face the problem of recovery of {\em vector-valued}
(instead of scalar-valued) components and our definition is adapted to
handle this.

Since the $\P_j$'s are projection matrices, we can also rewrite the
definition of fusion coherence as
\[
\mu_f = \max_{j \neq k} \left[ \absip{\a_j}{\a_k} \cdot
|\lambda_{\max}(\P_j\P_k)|^{1/2}\right]
\]
with $\lambda_{\max}$ denoting the largest eigenvalue, simply due to
the fact that the eigenvalues of $\P_k\P_j\P_k$ and $\P_j \P_k$
coincide. Indeed, if $\lambda$ is an eigenvalue of $\P_j \P_k$ with corresponding eigenvector $v$ then $\P_k \P_j \P_k v = \lambda \P_k v$. Since $\P_k^2 = \P_k$ this implies that $\P_k \P_j \P_k (\P_k v) = \lambda \P_k v$ so that $\lambda$ is an eigenvalue of $\P_k \P_j \P_k$ with eigenvector $\P_k v$. Let us also remark that $|\lambda_{\max}(\P_j\P_k)|^{1/2}$
equals the largest absolute value of the cosines of the principle
angles between $\cW_j$ and $\cW_k$.

Before we continue with the proof of the recovery condition in
Theorem~\ref{theo:main}, let us for a moment consider the following
special cases of this theorem.

\paragraph{Case $M=1$} In this case the projection matrices equal
$1$, and hence the problem reduces to the classical recovery problem
$\A\x=\y$ with $\x \in \RR^{N}$ and $\y \in \RR^n$. Thus our result
reduces to the result obtained in \cite{DE03}, and the fusion
coherence coincides with the commonly used mutual coherence, i.e.,
$\mu_f = \max_{j \neq k} \absip{\a_j}{\a_k}$.

\paragraph{Case $\cW_j = \RR^M$ for all $j$} In this case the problem
becomes the standard joint sparsity recovery. We recover a matrix
$\X^0 \in \RR^{N \times M}$ with few non-zero rows from knowledge of
$\A\X^0 \in \RR^{n \times M}$, without any constraints on the
structure of each row of $\X^0$. The recovered matrix is the fusion
frame representation of the sparse vector and each row $j$ represents
a signal in the subspace $\cW_j$ (the general case has the constraint
that $\X^0$ is required to be of the form $\U(\c^0)$). Again fusion
coherence coincides with the commonly used mutual coherence, i.e.,
$\mu_f = \max_{j \neq k} \absip{\a_j}{\a_k}$. 

\paragraph{Case $\cW_j \perp \cW_k$ for all $j,k$} In this case the fusion
coherence becomes 0. And this is also the correct answer, since in
this case there exists precisely one solution of the system
$\A\U(\c)=\Y$ for a given $\Y$.  Hence Condition
\eqref{eq:sparsitycondition} becomes meaningless.

\paragraph{General Case} In the general case we can consider two
scenarios: either we are given the subspaces $(\cW_j)_j$ or we are
given the measuring matrix $\A$.  In the first situation we face the
task of choosing the measuring matrix such that $\mu_f$ is as small
as possible. Intuitively, we would choose the vectors $(\a_j)_j$ so
that a pair $(\a_j,\a_k)$ has a large angle if the associated two
subspaces $(\cW_j,\cW_k)$ have a small angle, hence balancing the
two factors and try to reduce the maximum.
In the second situation, we can use a similar strategy now designing
the subspaces $(\cW_j)_j$ accordingly.\\

For the proof of Theorem \ref{theo:main} we
first derive a reformulation of the equation $\A\U(\c)=\Y$. 
For this, let $\P_j$ denote the orthogonal projection onto $\cW_j$ for each
$j=1,\ldots,N$, set
$\A_\P$ as in \eqref{eq:definitionAP}
and define the map $\varphi_k : \RR^{k \times M} \to \RR^{kM}$, $k
\ge 1$ by
\[
\varphi_k(\Z) = \varphi_k\left( \begin{array}{c}
\z_1\\
\hline\vspace*{-0.5cm} \\
\vdots\\
 \hline\vspace*{-0.45cm}\\
\z_k
\end{array} \right)
= (\z_1 \ldots \z_k)^T,
\]
i.e., the concatenation of the rows. 
Then it is easy to see that
\beq \label{eq:reformulation} \A \U(\c)
= \Y \quad \Leftrightarrow \quad \A_\P \varphi_N(\U(\c)) =
\varphi_n(\Y).
\eeq

We now split the proof of Theorem \ref{theo:main} into two lemmas,
and wish to remark that many parts are closely inspired by the
techniques employed in \cite{DE03,grni03}.
We first show that $\c^0$ satisfying \eqref{eq:sparsitycondition} is
the {\em unique} solution of $(P_1)$.

\begin{lemma}
\label{lem:l1uniqueness} If there exists a solution $\U(\c^0) \in
\RR^{N \times M}$ of the system $\A\U(\c)=\Y$ with $\c^0$ satisfying
\eqref{eq:sparsitycondition}, then $\c^0$ is the unique solution of
$(P_1)$.
\end{lemma}
\begin{IEEEproof}
We aim at showing that the condition on the fusion coherence implies
the fusion NSP. To this end, let $\h \in {\cal{N}} \setminus \{0\}$, i.e.,
$\A \U(\h) = 0$. By using the reformulation
\eqref{eq:reformulation}, it follows that
\[
\A_\P \varphi_N(\U(\h)) = 0.
\]
This implies that
\[
\A_\P^* \A_\P \varphi_N(\U(\h)) = 0.
\]
Defining $\a_j$ by $\a_j = (a_{ij})_i$ for each $j$, the previous
equality can be computed to be
\[
(\ip{\a_j}{\a_k} \P_j\P_k)_{jk} \varphi_N(\U(\h)) = 0.
\]
Recall that we have required the vectors $\a_j$ to be normalized.
Hence, for each $j$,
\[
\U_j\h_j = - \sum_{k \neq j} \ip{\a_j}{\a_k} \P_j\P_k\U_k\h_k.
\]
Since $\norm{\U_j\h_j}_2 = \norm{\h_j}_2$ for any $j$, this gives
\begin{eqnarray*}
\norm{\h_j}_2 &\le & \sum_{k \neq j} \absip{\a_j}{\a_k} \cdot
\norm{\P_j\P_k}_{2 \to 2} \norm{\h_k}_2 \\
& \le & \mu_f (\norm{\h}_{2,1} -
\norm{\h_j}_2),
\end{eqnarray*}
which implies
\[
\norm{\h_j}_2 \le (1+\mu_f^{-1})^{-1} \norm{\h}_{2,1}.
\]
Thus, we have
\begin{eqnarray*}
\norm{\h_S}_{2,1} & \le & \#(S) \cdot (1+\mu_f^{-1})^{-1}\norm{\h}_{2,1}\\
& = & \norm{\c^0}_{2,0} \cdot (1+\mu_f^{-1})^{-1}\norm{\h}_{2,1}.
\end{eqnarray*}
Concluding, \eqref{eq:sparsitycondition} and the fusion null space
property show that $\h$ satisfies \eqref{eq:inequality_h} unless
$\h=0$, which implies that $\c^0$ is the unique minimizer of $(P_1)$
as claimed.
\end{IEEEproof}

Using Lemma~\ref{lem:l1uniqueness} it is easy to show the following
lemma.
\begin{lemma}
\label{lem:l0uniqueness} If there exists a solution $\U(\c^0) \in
\RR^{N \times M}$ of the system $\A\U(\c)=\Y$ with $\c^0$ satisfying
\eqref{eq:sparsitycondition}, then $\c^0$ is the unique solution of
$(P_0)$.
\end{lemma}

\begin{IEEEproof} Assume $\c^0$ satisfies \eqref{eq:sparsitycondition} and $\A\U(\c^0)=\Y$.
Then, by Lemma \ref{lem:l1uniqueness}, it is the unique solution of
$(P_1)$. Assume there is a $\tilde{\c}$ satisfying $\A
\U(\tilde{\c})=\Y$ such that $\|\tilde{\c}\|_{2,0} \leq \|\c^0\|_{2,0}$. Then
$\tilde{\c}$ also satisfies \eqref{eq:sparsitycondition} and again
by Lemma \ref{lem:l1uniqueness} $\tilde{\c}$ is also the unique
solution to $(P_1)$. But this means that $\tilde{\c} = \c^0$ and
$\c^0$ is the unique solution to $(P_0)$.
\end{IEEEproof}

We observe that Theorem \ref{theo:main} now follows immediately from
Lemmas \ref{lem:l1uniqueness} and \ref{lem:l0uniqueness}.

\subsection{Fusion Restricted Isometry Property}
\label{sec:RIP_recovery}
Finally, we consider the condition for sparse recovery using the
restricted isometry property (RIP) of the sampling matrix.  The RIP
property on the sampling matrix, first introduced in~\cite{CRT06b},
complements the null space propery and the mutual coherence
conditions. Definition~\ref{def:frip} generalizes it for the fusion
frame setup.  Informally, we say that $(\A, (\cW_j)_{j=1}^N)$
satisfies the {\em fusion restricted isometry property} (FRIP) if
$\delta_k$ is small for reasonably large $k$. Note that we obtain the
classical definition of the RIP of $\A$ if $M=1$ and all the subspaces
$\cW_j$ have dimension $1$. Using this property we can prove
Theorem~\ref{theo_FRIP1}.

\begin{IEEEproof}[Proof of Theorem~\ref{theo_FRIP1}] The proof proceeds analogously to the one of Theorem 2.6 in \cite{ra09}, that
is, we establish the fusion NSP. The claim will then
follow from Theorem~\ref{lem:FNSP}.

Let us first note that
\[
|\langle \A_\P \u, \A_\P \vv \rangle| \leq \delta_k \|\u\|_{2,2} \|\vv\|_{2,2}
\]
for all $\u = (\u_1,\hdots,\u_N),\vv=(\vv_1,\hdots,\vv_N) \in \RR^{MN}$, $\u_j,\vv_j \in \RR^M$, with
$\supp\, \u = \{j: \u_j \neq 0\} \cap \supp\, \vv = \emptyset$ and $\|\u\|_{2,0}  + \|\vv\|_{2,0} \leq k$.
This statement follows completely analogously to the proof of
Proposition 2.5(c) in \cite{ra09-1}, see also
\cite{ca08,cdd09}.

Now let $\h \in {\cal{N}} = \{\h: \A \U(\h) = 0\}$ be given. Using the reformulation \eqref{eq:reformulation}, it follows that
\[
\A_\P \varphi_N(\U(\h)) = 0.
\]
In order to show the fusion NSP it is enough to consider an index set $S_0$ of size $k$
of largest components $\|\h_j\|_2$, i.e., $\|\h_j\|_2 \geq \|\h_i\|_2$ for all $j \in S_0$,
$i \in S_0^c =\{1,\hdots,N\} \setminus S_0$. We partition $S_0^c$ into index sets
$S_1, S_2,\hdots$ of size $k$ (except possibly the last one), such that $S_1$ is an index set
of largest components in $S_0^c$, $S_2$ is an index set of largest components in $(S_0 \cup S_0)^c$, etc. Let $\h_{S_i}$ be the vector that coincides with $\h$ on $S_i$ and is set to zero
outside. In view of $\h \in {\cal{N}}$ we have
$\A \U(\h_{S_0}) = \A \U(-\h_{S_1} - \h_{S_2} - \cdots)$.
Now set $\z = \varphi_N(\U(\h))$ and $\z_{S_i} = \varphi_N(\U(\h_{S_i}))$. It follows
that $\A_\P(\z_{S_0}) = \A_\P(-\sum_{i \geq 1} \z_{S_i})$.
By definition of the FRIP we obtain
\begin{eqnarray*}
\|\h_{S_0}\|_{2,2}^2 &=& \|\z_{S_0}\|_{2,2}^2 \leq \frac{1}{1-\delta_{k}} \|\A_\P \z_{S_0} \|_{2,2}^2\\
& = & \frac{1}{1-\delta_{k}} \Big\langle \A_\P \z_{S_0}, \A_\P\big(-\sum_{i\geq 1} \z_{S_i}\big) \Big\rangle\\
&\leq&  \frac{1}{1-\delta_k} \sum_{i\geq 1} |\langle \A_\P \z_{S_0}, \A_\P(-\z_{S_i}) \rangle|\\
&\leq& \frac{\delta_{2k}}{1-\delta_k} \|\z_{S_0}\|_{2,2} \cdot\sum_{i \geq 1} \|\z_{S_i}\|_{2,2}.
\end{eqnarray*}
Using that $\delta_k \leq \delta_{2k}$ and dividing by $\|\z_{S_0}\|_{2,2}$ yields
\[
\|\z_{S_0}\|_{2,2} \leq \frac{\delta_{2k}}{1-\delta_{2k}} \sum_{i \geq 1} \|\z_{S_i}\|_{2,2}.
\]
By construction of the sets $S_i$ we have
$
\|\z_j\|_2 \leq \frac{1}{k} \sum_{\ell \in S_{i-1}} \|\z_\ell\|_2 
$
for all $j \in S_i$, hence,
\[
\|\z_{S_i}\|_{2,2} = \big( \sum_{j \in S_i} \|\z_j\|_2^2 \big)^{1/2} \leq \frac{1}{\sqrt{k}} \|\z_{S_{i-1}}\|_{2,1}.
\]
The Cauchy-Schwarz inequality yields
\begin{eqnarray*}
\|\h_{S_0}\|_{2,1} & \leq & \sqrt{k} \|\h_{S_0}\|_{2,2} \leq \frac{\delta_{2k}}{1-\delta_{2k}}
\sum_{i\geq 1} \|\z_{S_{i-1}}\|_{2,1}\\
& \leq & \frac{\delta_{2k}}{1-\delta_{2k}} (\|\z_{S_0}\|_{2,1} + \|\z_{S_0^c}\|_{2,1}) < \frac{1}{2} \|\h\|_{2,1},
\end{eqnarray*}
where we used the assumption $\delta_{2k} < 1/3$. Hence, the fusion null space
property follows.
\end{IEEEproof}

Having proved that the FRIP ensures signal recovery, our next
proposition relates the classical RIP with our newly introduced
FRIP. Let us note, however, that using the RIP to guarantee the FRIP
does not take into account any properties of the fusion frame, so it
is sub-optimal---especially if the subspaces of the fusion frame are
orthogonal or almost orthogonal.

\begin{proposition} Let $\A \in \RR^{n \times N}$ with classical
restricted isometry constant $\tilde{\delta}_k$, that is,
\[
(1-\tilde{\delta}_k) \|y\|_2^2 \leq \|\A y\|_2^2 \leq
(1+\tilde{\delta}_k) \|y\|_2^2
\]
for all $k$-sparse $y \in \RR^N$. Let $(\cW_j)_{j=1}^N$ be an arbitrary
fusion frame for $\RR^M$.  Then the fusion restricted isometry constant
$\delta_k$ of $(\A, (\cW)_{j=1}^N)$ satisfies $\delta_k \leq
\tilde{\delta}_k$.
\end{proposition}

\begin{IEEEproof}
Let $\c$ satisfy $\norm{\c}_0 \le k$, and denote the columns of the matrix $\U(\c)$ by
$\u_1,\ldots,\u_M$. The condition $\norm{\c}_0 \le k$ implies that each $\u_i$ is $k$-sparse.
Since $\A$ satisfies the RIP of order $k$ with constant $\delta_k$, we obtain
\begin{eqnarray*}
\norm{\A \U(\c)}_{2,2}^2 & =  & \sum_{i=1}^M \norm{\A \u_i}_2^2 \le (1+\delta_k) \sum_{i=1}^M \norm{\u_i}_2^2\\
&= &(1+\delta_k) \norm{\U(\c)}_{2,2}^2 
 =  (1+\delta_k) \norm{\c}_{2,2}^2
\end{eqnarray*}
as well as
\begin{eqnarray*}
\norm{\A \U(\c)}_{2,2}^2 & = & \sum_{i=1}^M \norm{\A \u_i}_2^2 \ge (1-\delta_k) \sum_{i=1}^M \norm{\u_i}_2^2\\
& = & (1-\delta_k) \norm{\U(\c)}_{2,2}^2 = (1-\delta_k) \norm{\c}_{2,2}^2.
\end{eqnarray*}
This proves the proposition because $\A \U(\c) = \A_\P \U(\c)$.
\end{IEEEproof}

\subsection{Additional Remarks and Extensions}
\label{RIP_Remarks}

Of course, it is possible to extend the proof of Theorem~\ref{theo_FRIP1} in a similar manner
to \cite{CRT06b,ca08,fola09,fo09} such that we can accommodate measurement noise and
signals that are well approximated by sparse fusion frame
representation. We state the analog of the main theorem of \cite{ca08} 
without proof.

\begin{theorem} Assume that the fusion restricted isometry constant $\delta_{2k}$ of
$(\A,(\cW_j)_{j=1}^N)$ satisfies
\[
\delta_{2k} <  
\Delta := \sqrt{2}-1 \approx 0.4142.
\]
For $\x \in \cH$, let noisy measurements
$\Y = \A \x + \eta$ be given with $\|\eta\|_2 \leq \epsilon$. Let $\c^\#$ be the solution
of the convex optimization problem
\[
\min \|\c\|_{2,1} \quad \mbox{subject to} \quad \|\A \U(\c) - \Y\|_{2,2} \leq \eta,
\]
and set $\x^\# = \U(\c^\#)$. Then
\[
\|\x - \x^\#\|_{2,2} \leq C_1 \eta + C_2 \frac{\|\x^k - \x\|_{2,1}}{\sqrt{k}}
\]
where $\x_k$ is obtained from $\x$ be setting to zero all components except
the $k$ largest in norm. The constants $C_1,C_2 > 0$ only depend on $\delta_{2k}$ (or rather
on $\Delta - \delta_{2k}$).
\end{theorem}

\section{Probabilistic Analysis}
\label{sec:average_case}

\subsection{General Recovery Condition}

We start our analysis by deriving a recovery condition on the
measurement matrix and the signal which the reader might want to
compare with \cite{fu04,tr05-1}. Given a matrix $\X \in \RR^{N \times
  M}$, we let ${\rm sgn}(\X) \in \RR^{N \times M}$ denote the matrix
which is generated from $\X$ by normalizing each entry $X_{ji}$ by the
norm of the corresponding row $\X_{j,\cdot}$.  More precisely,
\[
{\rm sgn}(\X)_{ji} = \left\{ \begin{array}{ccl}
  \frac{X_{ji}}{\norm{\X_{j,\cdot}}_2} & \mbox{~if~} &
  \norm{\X_{j,\cdot}}_2 \neq 0,\\ 0 & \mbox{~if~} &
  \norm{\X_{j,\cdot}}_2=0.
\end{array} \right.
\]
Column vectors are defined similarly by $\X_{\cdot,i}$.

Under a certain condition on $A$, which is dependent on the support
of the solution, we derive the result below on unique recovery. To
phrase it, let $(\cW_j)_{j=1}^N$ be a fusion frame with associated
orthogonal bases $(\U_j)_{j=1}^N$ and orthogonal projections
$(\P_j)_{j=1}^N$, and recall the definition of the notion $\U(\c)$
in Section \ref{sec:formulation}.
Then, for some support set $S = \{j_1, \ldots, j_{|S|}\}
\subseteq \{1,\ldots,N\}$ of $\U(\c)$,
 we let
\[
\A_S = (\A_{\cdot,j_1} \cdots \A_{\cdot,j_{|S|}}) \in \RR^{n \times
|S|}
\]
and
\[
\U(\c)_S = \left( \begin{array}{c}
\c_{j_1}^T \U_{j_1}^T\\
\hline\vspace*{-0.5cm} \\
\vdots\\
 \hline\vspace*{-0.45cm}\\
\c_{j_{|S|}}^T \U_{j_{|S|}}^T
\end{array} \right)
\in  \RR^{|S| \times M}.
\]

Before stating the theorem, we wish to remark that its proof uses similar
ideas as the analog proof in \cite{elra09}. We however state all details
for the convenience of the reader.

\begin{theorem}
\label{theo:conditionH} Retaining the notions from the beginning of this section,
we let $\c_j \in \RR^{m_j}$, $j = 1 \ldots, N$
with $S = \supp(\c) = \{j : \c_j \neq 0\}$. If $\A_S$ is non-singular and there exists
a matrix $\H \in \RR^{n \times M}$ such that
\beq
\label{eq:conditionH1} \A_S^T \H = {\rm sgn}(\U(\c)_S) \eeq and \beq
\label{eq:conditionH2} \norm{\H^T \A_{\cdot,j}}_2 < 1 \quad
\mbox{for all } j \not\in S,
\eeq
then $\U(\c)$ is the unique solution of $(P_1)$.
\end{theorem}

\begin{IEEEproof}
Let $\c_j \in \RR^{m_j}$, $j = 1 \ldots, N$ be a solution of $\Y =
\A \U(\c)$, set $S = \supp(\U(\c))$, and suppose $\A_S$ is
non-singular and the hypotheses \eqref{eq:conditionH1} and
\eqref{eq:conditionH2} are satisfied for some matrix $\H \in \RR^{n
\times M}$. Let $\c'_j \in \RR^{m_j}$, $j = 1 \ldots, N$ with
$\c'_{j_0} \neq \c_{j_0}$ for some $j_0$ be a different set of
coefficient vectors which satisfies
$\Y = \A \U(\c')$. To prove our result we aim to establish that 
\beq \label{eq:claimconditionH} \norm{\c}_{2,1} < \norm{\c'}_{2,1}.
\eeq
We first observe that
\begin{eqnarray*}
\norm{\c}_{2,1} & = & \norm{\U(\c)}_{2,1} = \norm{\U(\c)_S}_{2,1}\\ 
&=&
\tr\left[{\rm sgn}(\U(\c)_S)(\U(\c)_S)^T\right].
\end{eqnarray*}
Set $S' = \supp(\U(\c'))$, apply \eqref{eq:conditionH1}, and exploit
properties of the trace,
\begin{eqnarray*} 
\label{eq:conditionH_eq1}
\norm{\c}_{2,1} &=& \tr\left[\A_S^T \H (\U(\c)_S)^T\right] =
\tr\left[(\A \U(\c))^T \H\right] \\
&=& \tr\left[(\A \U(\c'))^T \H\right]\\
& = & \tr\left[(\A \U(\c'_S))^T \H\right] + \tr\left[(\A \U(\c'_{S^c}))^T \H\right].
\end{eqnarray*}
Now use the
Cauchy-Schwarz inequality to obtain
\begin{eqnarray*}
\norm{\c}_{2,1} & \leq & \sum_{j \in S} \norm{(\U(\c')_{S})_{j,\cdot}}_2
\norm{(\H^T \A_{S})_{\cdot,j}}_2\\
 &+&  \sum_{j \in S^c} \norm{(\U(\c')_{S^c})_{j,\cdot}}_2
\norm{(\H^T \A_{S^c})_{\cdot,j}}_2 \\
& \leq & \max_{j \in S} \norm{(\H^T \A_{S})_{\cdot,j}}_2  \norm{\c'_{S}}_{2,1} \\
&+& \max_{j \in S^c} \norm{(\H^T \A_{S^c})_{\cdot,j}}_2  \norm{\c'_{S^c}}_{2,1}\\
  &<& \|\c'_S\|_{2,1} + \|\c'_{S^c}\|_{2,1} = \|\c'\|_{2,1}.
\end{eqnarray*}
The strict inequality follows from $\|\c'_{S^c}\|_1 > 0$, which is true because
otherwise $\c'$ would be supported on $S$. The equality $\A \U(\c) = \A \U(\c')$
would then be in contradiction to the injectivity of $\A_S$ (recall that $\c \neq \c'$).
This concludes the proof.
\end{IEEEproof}

The matrix $\H$ exploited in Theorem \ref{theo:conditionH} might be
chosen as
\[
\H = (\A_S^\dagger)^T {\rm sgn}(\U(\c)_S)
\]
to satisfy \eqref{eq:conditionH1}. This particular choice will in
fact be instrumental for the average case result we are aiming for.
For now, we obtain the following result as a corollary from Theorem
\ref{theo:conditionH}.

\begin{corollary}
\label{coro:recoverycondition} Retaining the notions from the beginning of this section,
we let $\c_j \in \RR^{m_j}$, $j = 1
\ldots, N$, with $S = \supp(\U(\c))$. If $\A_S$ is non-singular and
\beq \label{eq:conditiononA} \norm{{\rm sgn}(\U(\c)_S)^T
\A_S^\dagger \A_{\cdot,j}}_2 < 1 \quad \mbox{for all } j \not\in S,
\eeq then $\U(\c)$ is the unique solution of $(P_1)$.
\end{corollary}

For later use, we will introduce the matrices $\tilde{\U}_j \in
\RR^{M \times Nm}$ defined by
\[
\tilde{\U}_j = ({\bf 0}_{M \times m} | \cdots | {\bf 0}_{M \times m}
| \U_j | {\bf 0}_{M \times m} | \cdots | {\bf 0}_{M \times m}),
\]
where $\U_j$ is the $j$th block. 
For some
$\b = (b_1,\ldots,b_k)^T \in \RR^k$, we can then write $\U(\c)_S^T
\b$ as
\begin{equation}\label{def:Uc}
\sum_{\ell=1}^k b_\ell\tilde{\U}_{j_\ell} \X \in \RR^M.
\end{equation}

Using the above and the probabilistic model in
Sec.~\ref{sec:prob_result_intro} to prove our main probabilistic
result, Theorem~\ref{thm:main}.

\subsection{Probability of Sparse Recovery for Fusion Frames}

The proof of our probabilistic result, Theorem~\ref{thm:main}, is
developed in several steps. A key ingredient is a concentration of
measure result: If $f$ is a Lipschitz function on $\RR^K$ with
Lipschitz constant $L$, i.e., $|f(x)-f(y)| \leq L \|x-y\|_2$ for all
$x,y \in \RR^K$, $\X$ is a $K$-dimensional vector of independent standard normal random
variables then \cite[eq. (2.35)]{le01}
\begin{equation}\label{conc:ineq}
\PP(|f(X) - \EE f(X)| \geq u) \leq 2e^{-u^2/(2L^2)} \quad \mbox{ for
all } u > 0.
\end{equation}

Our first lemma investigates the properties of a function related to
\eqref{def:Uc} that are needed to apply the above inequality.

\begin{lemma}
\label{lemma:helpprob2} Let $\b = (b_1,\ldots,b_k)^T \in \RR^k$ and
$S = \{j_1, \ldots, j_{k}\} \subseteq \{1,\ldots,N\}$.
Define the function $f$ by
\[
f(\X) = \norm{\sum_{\ell=1}^k b_\ell \tilde{\U}_{j_\ell} \X}_2,
\quad \X \in \RR^{Nm}.
\]
Then the following holds. \bitem
\item[{\rm (i)}] $f$ is Lipschitz with constant $\norm{\sum_{\ell=1}^k b_\ell \tilde{\U}_{j_\ell}}_{2 \to 2}$.
\item[{\rm (ii)}] For a standard Gaussian vector $\X \in \RR^{Nm}$
we have $\mathbb{E}[f(\X)] \le \sqrt{m} \norm{\b}_2$. \eitem
\end{lemma}

\begin{IEEEproof}
The claim in (i) follows immediately from
\begin{eqnarray*}
|f(\X)-f(\Y)|
 &=&  \Big|\norm{\sum_{\ell=1}^k b_\ell \tilde{\U}_{j_\ell} \X}_2 - \norm{\sum_{\ell=1}^k b_\ell \tilde{\U}_{j_\ell} \Y}_2\Big|\\
& \le & \norm{\left(\sum_{\ell=1}^k b_\ell
\tilde{\U}_{j_\ell}\right)(\X-\Y)}_2\\
&\le& \norm{\sum_{\ell=1}^k b_\ell \tilde{\U}_{j_\ell}}_{2 \to 2}
\norm{\X-\Y}_2.
\end{eqnarray*}
It remains to prove (ii). Obviously,
\begin{eqnarray*}
\left(\mathbb{E} f(\X)\right)^2 & \le & \mathbb{E}[f(\X)^2] =
\mathbb{E}\left[\sum_{i=1}^M \Big|\sum_{\ell=1}^k b_\ell
(\tilde{\U}_{j_\ell} \X)_i \Big|^2\right] \\
&=& \sum_{i=1}^M
\sum_{\ell,\ell'=1}^k b_\ell b_{\ell'}
\mathbb{E}[(\tilde{\U}_{j_\ell} \X)_i (\tilde{\U}_{j_{\ell'}}
\X)_i].
\end{eqnarray*}
Invoking the conditions on $\X$,
\begin{eqnarray*}
\mathbb{E}[f(\X)]^2
 &\le& \sum_{i=1}^M \sum_{\ell=1}^k b_\ell^2 \mathbb{E}[(\tilde{\U}_{j_\ell} \X)_i]^2
 = \sum_{\ell=1}^k b_\ell^2 \norm{\tilde{\U}_{j_\ell}}_F^2\\
& =& m \norm{\b}_2^2.
\end{eqnarray*}
\end{IEEEproof}

Next we estimate the Lipschitz constant of the function $f$ in the
previous lemma.

\begin{lemma}
\label{lemma:helpprob1} Let $\b = (b_1,\ldots,b_k)^T \in \RR^k$ and
$S = \{j_1, \ldots, j_{k}\} \subseteq \{1,\ldots,N\}$. Then
\[
\norm{\sum_{\ell=1}^k b_\ell \tilde{\U}_{j_\ell}}_{2 \to 2} \le
\norm{\b}_\infty \sqrt{1+\max_{i \in S} \sum_{j \in S, j \neq i}
\lambda_{\max{}}(\P_i \P_j)^{1/2}}.
\]
\end{lemma}

\begin{IEEEproof}
First observe that
\begin{eqnarray*}
\norm{\sum_{\ell=1}^k b_\ell \tilde{\U}_{j_\ell}}_{2 \to 2} 
&= &
\norm{(\sum_{\ell=1}^k b_\ell \tilde{\U}_{j_\ell})^T(\sum_{\ell=1}^k
b_\ell \tilde{\U}_{j_\ell})}_{2 \to 2}^{1/2} \\
& = & \norm{\sum_{\ell,
\ell'=1}^k b_\ell b_{\ell'} \tilde{\U}_{j_\ell}^T
\tilde{\U}_{j_{\ell'}}}_{2 \to 2}^{1/2}.
\end{eqnarray*}
Since
\[
\norm{\sum_{\ell, \ell'=1}^k b_\ell b_{\ell'} \tilde{\U}_{j_\ell}^T
\tilde{\U}_{j_{\ell'}}}_{2 \to 2} \le \norm{\b}_\infty^2
\norm{(\U_i^T \U_j)_{i,j \in S}}_{2 \to 2},
\]
it follows that
\beq \label{eq:lemmahelpprob1_eq1}
\norm{\sum_{\ell=1}^k b_\ell \tilde{\U}_{j_\ell}}_{2 \to 2} \le
\norm{\b}_\infty \norm{(\U_i^T \U_j)_{i,j \in S}}_{2 \to 2}^{1/2}.
\eeq
Next,
\begin{eqnarray}
\label{eq:lemmahelpprob1_eq2} 
&&\norm{(\U_i^T \U_j)_{i,j \in S}}_{2
\to 2} \le \max_{i\in S} \sum_{j\in S} \norm{\U_i^T \U_j}_{2 \to 2}\\
& = & 1 + \max_{i\in S} \sum_{j \in S, j \neq i} \norm{\U_i^T \U_j}_{2
\to 2}. \notag
\end{eqnarray} 
By definition of the orthogonal projections $\P_i$,
\begin{eqnarray*}
\norm{\U_i^T \U_j}_{2 \to 2}
 & = & \norm{(\U_i^T \U_j)^T \U_i^T \U_j}_{2 \to 2}^{1/2}\\
 & = & \norm{\U_i^T \U_j \U_j^T \U_i}_{2 \to 2}^{1/2}
 = \norm{\P_i \P_j}_{2 \to 2}^{1/2}\\
& = &\lambda_{\max{}}(\P_i \P_j)^{1/2}.
\end{eqnarray*}
Combining with \eqref{eq:lemmahelpprob1_eq2}, \beq
\label{eq:lemmahelpprob1_eq3} 
\norm{(\U_i^T \U_j)_{i,j \in S}}_{2
\to 2} \le 1 + \max_{i\in S} \sum_{j \in S, j \neq
i}\lambda_{\max{}}(\P_i \P_j)^{1/2}. \eeq The lemma now follows from
\eqref{eq:lemmahelpprob1_eq1}, \eqref{eq:lemmahelpprob1_eq2}, and
\eqref{eq:lemmahelpprob1_eq3}.
\end{IEEEproof}

Now we have collected all ingredients to prove our main result, Theorem~\ref{thm:main}

\begin{IEEEproof}
  Denote $\b^{(j)} = (b_1^{(j)},\ldots,b_k^{(j)})^T = \A_S^\dagger
  \A_{\cdot,j} \in \RR^k$ for all $j \not\in S$ and choose $\delta \in
  (0,1-\alpha^2)$. By Corollary \ref{coro:recoverycondition}, the
  probability that the minimization problem ($P_1$) fails to recover $\U(\c)$ from
  $\Y = \A \U(\c)$ can be estimated as
  \begin{eqnarray*}
    & &\PP \left(\max_{j \notin S} \norm{{\rm sgn}(\U(\c)_S)^T
      \b^{(j)}}_2 > 1\right)\\
    & = & \PP \left( \max_{j \not\in S} \norm{\sum_{\ell=1}^k b_\ell^{(j)} \norm{U_{j_\ell} \X_\ell}_2^{-1} \tilde{\U}_{j_\ell} \X}_2 > 1\right)\\
    & \leq &  \PP \left( \max_{j \not\in S} \norm{\sum_{\ell=1}^k b_\ell^{(j)}  \tilde{\U}_{j_\ell} \X}_2 > \sqrt{(1-\delta) m}     \right) \\
    & + & \PP\left( \max_{\ell=1,\hdots,k} \norm{U_{j_\ell} \X_\ell}_2 < \sqrt{(1-\delta) m} \right)\\
    & \leq & \sum_{j\notin S} \PP\left( \norm{\sum_{\ell=1}^k
      b_\ell^{(j)}  \tilde{\U}_{j_\ell} \X}_2 > \sqrt{(1-\delta) m}
    \right) \\
    & + &\sum_{\ell=1}^k \PP \left(\norm{\X_\ell}_2^2 \leq (1-\delta) m
    \right).
  \end{eqnarray*}
  Since $\X_\ell$ is a standard Gaussian vector in $\RR^m$
  \cite[Corollary 3]{ba05} gives
  \[
  \PP \left(\norm{\X_\ell}_2^2 \leq (1-\delta) m \right) \leq \exp(-\delta^2m/4).
  \]
  Furthermore, the concentration inequality \eqref{conc:ineq} combined
  with Lemmas \ref{lemma:helpprob2} and Lemma \ref{lemma:helpprob1}
  yields
  \begin{eqnarray*}
    &&\PP\left( \norm{\sum_{\ell=1}^k b_\ell^{(j)}  \tilde{\U}_{j_\ell}
      \X}_2 > \sqrt{(1-\delta) m}
    \right)\\
    &=& \PP\left( \norm{\sum_{\ell=1}^k b_\ell^{(j)}  \tilde{\U}_{j_\ell} \X}_2 >  \|b^{(j)}\|_2 \sqrt{m} \right.\\
   && \phantom{ \PP\left(\sum_{j}^n \right)} \left. + (\sqrt{1-\delta} - \|b^{(j)}\|_2) \sqrt{m} \right)\\
     &\leq& \exp\left(-\frac{(\sqrt{1-\delta}-\|b^{(j)}\|_2)^2 m}{2 \|b^{(j)}\|_\infty^2 \theta}\right) \\
     &\leq& \exp\left(-\frac{(\sqrt{1-\delta}-\|b^{(j)}\|_2)^2 m}{2 \|b^{(j)}\|_2^2 \theta}\right) \\
    & \leq &  \exp\left(- \frac{(\sqrt{1-\delta}-\alpha)^2m}{2 \alpha^2
      \theta}\right).
  \end{eqnarray*}
  Combining the above estimates yields the statement of the Theorem.
\end{IEEEproof}

\section{Conclusions and Discussion}

The main contribution in this paper is the generalization of
standard Compressed Sensing results for sparse signals to signals
that have a sparse fusion frame representation. As we demonstrated,
the results generalize to fusion frames in a very nice and easy to
apply way, using the mixed $\ell_2/\ell_1$ norm.

A key result in our work shows that the structure in fusion frames
provides additional information that can be exploited in the
measurement process. Specifically, our definition of fusion
coherence demonstrates the importance of prior knowledge about the
signal structure. Indeed, if we know that the signal lies in
subspaces with very little overlap (i.e., where $\|\P_j\P_k\|_2$ is
small in Definition~\ref{def:fusion_coherence}) we can relax the
requirement on the coherence of the corresponding vectors in the
sampling matrix (i.e., $|\langle\a_j,\a_k\rangle|$ in the same
definition) and maintain a low fusion coherence. This behavior
emerges from the inherent structure of fusion frames.

The emergence of this behavior is evident both in the guarantees
provided by the fusion coherence, and in our average case
analysis. Unfortunately, our analysis of this property currently has
not been incorporated in a tight approach to satisfying the Fusion RIP
property, as described in Section~\ref{sec:RIP_recovery}.  While an
extension of such analysis for the RIP guarantees is desirable, it is
still an open problem.

Our average case analysis also demonstrates that as the sparsity
structure of the problem becomes more intricate, the worst case
analysis can become too pessimistic for many practical cases. The
average case analysis provides reassurance that typical behavior is as
expected; significantly better compared to the worst case. Our results
corroborate and extend similar findings for the special case of joint
sparsity in~\cite{elra09}.

\section*{Acknowledgement}
G. Kutyniok would like to thank Peter
Casazza, David Donoho, and Ali Pezeshki for inspiring discussions on
$\ell_1$ minimization and fusion frames. G. Kutyniok  would also
like to thank the Department of Statistics at Stanford University and
the Department of Mathematics at Yale University for their hospitality
and support during her visits.


\begin{IEEEbiographynophoto}{Petros T.\ Boufounos}
completed his undergraduate and graduate studies at MIT. He received the S.B.\ degree in Economics in 2000, the S.B.\ and M.Eng.\ degrees in Electrical Engineering and Computer Science (EECS) in 2002, and the Sc.D.\ degree in EECS in 2006. Since January 2009 he is with Mitsubishi Electric Research Laboratories (MERL) in Cambridge, MA.\ He is also a visiting scholar at the Rice University Electrical and Computer Engineering department.

Between September 2006 and December 2008, Dr.\ Boufounos was a postdoctoral associate with the Digital Signal Processing Group at Rice University doing research in Compressive Sensing. In addition to Compressive Sensing, his immediate research interests include signal acquisition and processing, data representations, frame theory, and machine learning applied to signal processing. He is also looking into how Compressed Sensing interacts with other fields that use sensing extensively, such as robotics and mechatronics. Dr.\ Boufounos has received the Ernst A.\ Guillemin Master Thesis Award  for his work on DNA sequencing and the Harold E.\ Hazen Award for  Teaching Excellence, both from the MIT EECS department. He has also  been an MIT Presidential Fellow. Dr.\ Boufounos is a member of the  IEEE, Sigma Xi, Eta Kappa Nu, and Phi Beta Kappa.
\end{IEEEbiographynophoto}

\begin{IEEEbiographynophoto}
{Gitta Kutyniok} received diploma degrees in mathematics and
computer science in 1996 and the Dr.~rer.~nat.\ in mathematics in 2000
from the University of Paderborn. In 2001, she spent one term as a
Visiting Assistant Professor at the Georgia Institute of Technology.
She then held positions as a Scientific Assistant at the University
of Paderborn and the University of Gie{\ss}en. From Oct. 2004 to
Sept. 2005, she was funded by a Research Fellowship by the
DFG-German Research Foundation and spent six months at each
Washington University in St. Louis and Georgia Institute of
Technology. She completed her Habilitation in mathematics in 2006
and received her venia legendi. In 2006, she was awarded a
Heisenberg Fellowship by the DFG. From Apr. 2007 to Sept. 2008, she
was a Visiting Fellow at Princeton University, Stanford University,
and Yale University. Since October 2008, she is a Professor for
Applied Analysis at the University of Osnabr\"uck. G.\ Kutyniok was
awarded the Weierstrass Prize for outstanding teaching of the
University of Paderborn in 1998, the Research Prize of the
University of Paderborn in 2003, the Prize of the University of
Gie{\ss}en in 2006, and the von Kaven Prize by the DFG in 2007. G.\
Kutyniok's research interests include applied harmonic analysis,
compressed sensing, frame theory, imaging sciences, and sparse
approximation.
\end{IEEEbiographynophoto}

\begin{IEEEbiographynophoto}
{Holger Rauhut} received the diploma degree in mathematics
from the Technical University of Munich
in 2001. He was a member of the graduate
program Applied Algorithmic Mathematics at the Technical University
of Munich from 2002 until 2004, and received the Dr.\ rer.\ nat.\ in mathematics
in 2004.
After his time as a
PostDoc in Wroc{\l}aw, Poland in 2005 (3 months), he has been with
the Numerical Harmonic Analysis Group at the Faculty of Mathematics of
the University of Vienna, where he completed has habilitation degree in 2008.
From March 2006 until February 2008
he was funded by an Individual
Marie Curie Fellowship from the European Union.
From March 2008, he is professor (Bonn Junior Fellow) at the Hausdorff Center for Mathematics
and Institute for Numerical Simulation at the University of Bonn. In 2010 he was awarded a
Starting Grant from the European Research Council.
H.\ Rauhut's research interests include
compressed sensing, sparse approximation,
time-frequency and wavelet analysis.
\end{IEEEbiographynophoto}

\end{document}